\documentclass{PoS}
\usepackage{enumerate}
\usepackage{enumitem}
\usepackage[subtle]{savetrees}

\title{EDGE: a code to calculate diffusion of cosmic-ray electrons and their gamma-ray emission}

\ShortTitle{EDGE}

\author{\speaker{R. Lopez-Coto}, J. Hahn, J.~Hinton, R. D. Parsons\\
        Max Planck Institut f\"ur Kernphysik, Heidelberg, Germany \\
        E-mail: \email{rlopez@mpi-hd.mpg.de}}

\author{F. Salesa Greus\\
        Instytut Fizyki Jadrowej im Henryka Niewodniczanskiego Polskiej Akademii Nauk, Krakow, Poland}

\author{S. BenZvi, M.U. Nisa\\
        Department of Physics \& Astronomy, University of Rochester, Rochester, NY, USA}

\author{H. Zhou\\
        Physics Division, Los Alamos National Laboratory, Los Alamos, NM 87545, USA}

\abstract{The positron excess measured by PAMELA and AMS can only be explained if there is one or several sources injecting them. Moreover, at the highest energies, it requires the presence of nearby ($\sim$hundreds of parsecs) and middle age (maximum of $\sim$hundreds of kyr) source. Pulsars, as factories of electrons and positrons, are one of the proposed candidates to explain the origin of this excess. To calculate the contribution of these sources to the electron and positron flux at the Earth, we developed EDGE (Electron Diffusion and Gamma rays to the Earth), a code to treat diffusion of electrons and compute their diffusion from a central source with a flexible injection spectrum. We can derive the source's gamma-ray spectrum, spatial extension, the all-electron density in space and the electron and positron flux reaching the Earth. We present in this contribution the fundamentals of the code and study how different parameters affect the gamma-ray spectrum of a source and the electron flux measured at the Earth.
}

\FullConference{35th International Cosmic Ray Conference, ICRC2017-\\
		10-20 July, 2017\\
		Bexco, Busan, Korea}

\begin{document}

\section{Introduction}
\label{sec:intro}

Cosmic rays (CRs) are high-energy charged particles that strike the atmosphere almost isotropically. They are composed by protons and helium nuclei (99\%), heavier nuclei, electrons (e$^-$), positrons (e$^+$), antiprotons and neutrinos. Currently, there are galactic propagation models that reproduce the e$^\pm$ CR energy spectrum assuming they are secondary products of the sea of CR collisions. 
However, the e$^+$ content in the total e$^\pm$ flux above 10 GeV (also known as ``positron excess'') measured by PAMELA and AMS amongst others, can only be explained if there is one or several sources injecting them. Moreover, taking into account standard diffusion and cooling of e$^\pm$, the extension of the e$^\pm$ spectrum up to TeV energies can only be explained if the source is nearby ($\sim$hundreds of parsecs) and middle age ($\sim$hundred of kyr). Pulsars, as factories of e$^\pm$, are one of the proposed candidates to explain the origin of this excess, however there are also more exotic explanations such as galactic jets \cite{Gupta2014} or dark matter \cite{Ibarra2013}. To measure the maximum energy reached by these e$^\pm$ at the Earth and unveil the origin of the positron excess is one of the most important questions unsolved in astroparticle physics nowadays. Since e$^\pm$ are charged particles, their arrival direction does not point to their origin because they are deflected by magnetic fields. One of the ways to study sources of CRs is to analyse the neutral subproducts of CR collisions such as gamma rays.

We developed EDGE (Electron Diffusion and Gamma rays to the Earth), a code to treat diffusion of electrons that was used in \cite{Geminga} to compute the electron diffusion from a central source, derive its $\gamma$-ray spectrum, profile and the all-electron density in the space. We present in this paper the fundaments of the code and study how different parameters affect the $\gamma$-ray spectrum of a source and the electron flux measured at the Earth.

\section{Diffusion of electrons}
\label{sec:diffusion}
If we assume a spherically symmetric case where electrons are diffusing from the pulsar, the equation that describes this process is:

\begin{equation}
\frac{\partial f}{\partial t} = \frac{D}{r^2}\frac{\partial}{\partial r}r^2\frac{\partial f}{\partial r}  + \frac{\partial}{\partial \gamma}(Pf) + Q
\end{equation}

where $\gamma$=E/m$_e$c$^2$ with E the energy of the particle, m$_e$ the mass of the electron and c the speed of light. $f(r,t,\gamma)$ is the energy distribution of particles at an instant t and distance r from the source, $D(\gamma$) the energy dependent diffusion coefficient, $P(\gamma$) the energy loss rate and $Q(r,t,\gamma$) is proportional to the injection spectrum. The details about how to solve this equation for particular cases can be found in  \cite{Atoyan95}. The Green function for this equation for an arbitrary injection spectrum $\Delta N (\gamma)$ is given by:

\begin{equation}
\label{eq:energy_density}
 f(r,t,\gamma) = \frac{\Delta N (\gamma_t) P(\gamma_t)}{\pi^{3/2}P(\gamma) r_{\textrm{diff}}^3} \exp \left( - \frac{r^2}{r_{\textrm{diff}}^2} \right)
\end{equation}

where $\gamma_t$ corresponds to the initial energy of the particles. The diffusion radius ($r_{\textrm{diff}}$) represents the mean free path of e$^\pm$ of a given energy. It is given by:

\begin{equation}
\label{eq:rdiff}
r_{\textrm{diff}}=2\sqrt{\Delta u}
\end{equation}

and $\Delta u$:

\begin{equation}
\Delta u = \int^{\gamma_t}_{\gamma}{  D(x) dx / P(x) } 
\label{eq:lambda}
\end{equation}

is the integral over the particle history from an initial energy $\gamma_t$ to an energy $\gamma$. 

\subsection{Injection spectrum}

\subsubsection{Time-dependence}
\label{sec:time-dependence}

Pulsars are rotating neutron stars that produce periodic radiation by spinning their powerful magnetic field through space. They loss their rotational energy by emitting a wind of electron and positron pairs that diffuses away when these particles escape outside of the pulsar's magnetosphere. We assume that this emission is isotropic and the wind is composed by the same quantity of electrons and positrons. Let us talk about some properties of pulsars that are important to characterize their emission.

The luminosity of a pulsar is given by \cite{Gaensler06}:
\begin{equation}
\label{eq:luminosity}
L(t) = L_0\left(   1 +  \frac{t}{\tau}\right)^{-\frac{n+1}{n-1}}
\end{equation}

where $n$ is the braking index of the pulsar, $L_0$ is the initial luminosity and $\tau$ the characteristic time. We assume that the pulsar behaves as a dipole, therefore $n$=3. 
The characteristic age $\tau_{\rm{c}}$ of a pulsar is estimated from its period and period derivative and is given by:

\begin{equation}
\tau_{\rm{c}} = \frac{P}{2 \dot{P}}
\end{equation}

\subsubsection{Spectral shape}

If we assume that the spectrum of injected electrons is given by a power-law:

\begin{equation}
\label{eq:injection_spectrum}
\frac{dN}{dE}=Q(\gamma,t) = Q_0 \gamma^{-\alpha}
\end{equation}

where $Q_0$ is the initial injection rate and $\alpha$ the injection rate's index. The injection rate is related to the pulsar's luminosity by the equation:

\begin{equation}
L_e(t) = \int^{\gamma \rm{min}}_{{\gamma \rm{max}}}Q(\gamma,t) \gamma m_{\rm{e}} c^2 d\gamma
\end{equation}

with $L_e(t)=\mu L(t)$. $L(t)$ is given by equation \ref{eq:luminosity} and $\mu$ is a constant $<$1 that determines the fraction of the luminosity that is transferred to electrons. The initial injection rate is therefore given by:

\begin{equation}
Q_0 = \left( \int^{\gamma \rm{min}}_{{\gamma \rm{max}}}\gamma^{-\alpha} \gamma m_{\rm{e}} c^2 d\gamma \right)^{-1} \frac{1}{\mu L_0}\left(   1 +  \frac{t}{\tau}\right)^{2}
\end{equation}

\subsection{Energy loss}
The energy loss rate is given by: 

\begin{equation}
\label{eq:energyloss}
P(\gamma) = -\frac{d \gamma}{dt}
\end{equation}

here we include synchrotron, Inverse Compton (IC) and bremsstrahlung losses.

\subsubsection{Inverse Compton:} We calculate the IC cross-section in two different regimes, depending on the target photon ($E_{\rm{ph}}$) and electron ($E_e$) energies:
\label{sec:inverse_compton}
\begin{enumerate}

\item if ($E_{\rm{ph}} \gamma_e / (m_e c^2)  < 0.1 $) 

we use the Thomson equation for the IC losses:

\begin{equation}
P(\gamma)_{\rm{IC}} = \frac{4 \sigma_T c}{3} U_{\rm{ph}} \gamma^2
\end{equation}

with $\sigma_T$ the Thomson IC cross-section, $c$ the speed of light and $U_{\rm{ph}}$ the energy density of the target photons.

\item if ($E_{\rm{ph}} \gamma_e / (m_e c^2)  > 0.1 $) 

P($\gamma_{\rm{IC}}$) is calculated using the full Klein-Nishina equation \cite{Blumenthal70}. We use equation 2.48 of \cite{Blumenthal70} to calculate the emissivity of the radiation fields and integrate it over the photon and electron spectra. The radiation targets are added as a grey body distribution with mean the temperature of each of the radiation fields.
\end{enumerate}

\subsubsection{Synchrotron:}

The synchrotron losses are given by \cite{Moderski05}:

\begin{equation}
P(\gamma)_{\rm{syn}} = \frac{4 \sigma_T c}{3} \frac{B^2}{8\pi} \gamma^2
\end{equation}

with $B$ the magnetic field experienced by the particles.

\subsubsection{Bremsstrahlung:}

The bremsstrahlung energy losses are taken from Eq. 9 of \cite{Haug04} for $E> 500$ keV. The ambient density $n$ assumed is 1~cm$^{-3}$
 

\subsection{Diffusion coefficient}
The energy-dependent diffusion coefficient is given by:

\begin{equation}
\label{eq:diff_coefficient}
D(\gamma) = D_{0} \left(   1+ \frac{\gamma}{\gamma^{\star}}  \right)^\delta
\end{equation}

where D$_{0}$ is the diffusion coefficient normalization, $\gamma^{\star}$ is usually taken $\sim$ 6$\times$10$^3$ (corresponding to $E^{\star}$=3 GeV) and $\delta$ the diffusion exponent, usually between 0.3 and 0.6 \cite{Atoyan95,Yuksel08}.

\subsection{Electron/positron flux at the Earth}
\label{sec:electron_positron_flux}
The electron flux at the Earth for a given source with an age $t_{\rm{age}}$ and situated at a distance $d$ is given by:

\begin{equation}
\label{eq:flux_earth}
J(\gamma) = \frac{c}{4\pi} f(d_{\rm{Earth}},t_{\rm{age}},\gamma)
\end{equation}

where $d_{\rm{Earth}}$ is the distance from the source to the Earth and $t_{\rm{age}}$ the age of the source.

\subsection{Positron fraction at the Earth}
\label{sec:electron_positron_fraction}

The positron fraction for a given $\gamma$ will be given by:

\begin{equation}
\label{eq:fraction_earth}
\frac{\rm{e}^+}{\rm{e}^++\rm{e}^-} (\gamma)= \frac{0.5 \times \rm{J(\gamma) + Sec.[e}^+]}{\rm{J(\gamma)+ Sec.[e}^+]+\rm{Prim.[e}^-]+\rm{Sec.[e}^-]}
\end{equation}

where the primary and secondary electron and positron fluxes are coming from secondary CR interactions. For these fluxes, we use the theoretical from the right panel of Figure 5 on \cite{Moskalenko97}.

\section{Electron, positron and $\gamma$-ray fluxes calculations using EDGE}
\label{sec:edge}

The \textbf{E}lectron \textbf{D}iffusion and \textbf{G}amma rays at the \textbf{E}arth (\textbf{EDGE}) is a flexible code that accepts as input parameters different pulsar and environment characteristics and gives as a final output the $\gamma$-ray spectrum produced by the source and the electron and positron flux produced at the Earth by this source.  
The code uses dependencies several dependencies from the GAMERA package \cite{GAMERA}. We will give an overview of the calculations performed by the code and the results obtained with a set of selected parameters.

\subsection{Default paremeters}
In the following, we will give a brief description of the default parameters used in the code. 

\subsubsection{General parameters}
The age of the system is selected $t_{\rm{age}}=300$ kyr and the distance to the Earth $d_{\rm{Earth}}$=250 pc.

\subsubsection{Injection spectrum}

Since we would like to derive a $\gamma$-ray spectrum together with the all-electron spectrum at the Earth, we will use the pulsar-like injection mechanism described in Section \ref{sec:time-dependence}. 
The minimum and maximum energy of the simulated electron spectrum are selected $E_{\rm{min}}$=1 GeV and $E_{\rm{max}}$=500 TeV. The injection spectrum index assumed is $\alpha$=2.2. The fraction of spin-down power transformed into $\gamma$-ray emission assumed is $\mu$=0.5.

\subsubsection{Energy losses}
The target photon fields used are:
\begin{itemize}

\item CMB: $\epsilon_{\rm{CMB}}=0.26\ $eV/cm$^3$, T=2.7 K
\item Infrarred: $\epsilon_{\rm{IR}}=0.3\ $eV/cm$^3$, T=20 K
\item Optical: $\epsilon_{\rm{Opt}}=0.3\ $eV/cm$^3$, T=5000 K
\end{itemize}
The default value for the magnetic field is $B=3 \mu$G.

\subsubsection{Diffusion coefficient}
We select $\delta$=0.33 driven by \cite{Kolmogorov1941} and in agreement with recent measurements \cite{Aguilar2016_BCratio}.

\subsection{Energy density in space}
The procedure to get the electron and positron flux at the Earth starts by evaluating the energy density of the electrons produced by the central pulsar (Eq. \ref{eq:energy_density}) for a time $t=t_{\rm{age}}$ in every point of the space and for the full range of energies. If we numerically solve Eq. \ref{eq:lambda} and \ref{eq:energyloss} and given all the other parameters, we obtain a look-up table with the energy density of e$^\pm$ for different energies and at different distances from the pulsar. If we consider cooling in the Thomson regime, we can use the cooling time of electrons in the Thomson regime, given by:

\begin{equation}
\label{eq:cooling}
t_{\rm{cool}} \approx 3 \times 10^5 \left( \frac{E}{\rm{TeV}} \frac{\epsilon}{\rm{eV\ cm}^{-3}} \right)^{-1} \rm{yr}
\end{equation}

where $E$ the energy of the electrons and $\epsilon$ is the energy density of the target photons plus the magnetic field. The diffusion radius has therefore the following dependencies:

\begin{enumerate}[label=\roman*)]

\item if $t_{\rm{cool}} (E) < t_{\rm{age}}$ (corresponding to $E\gtrsim1$ TeV), the system is cooling-limited and the diffusion radius $r_{\rm{diff}}\propto E^{(\delta-1)/2}$ increases with decreasing energy.

\item if $t_{\rm{cool}} (E) > t_{\rm{age}}$ (corresponding to $E\lesssim1$ TeV), the system is age-limited and the diffusion radius is $r_{\rm{diff}}\propto E^{\delta/2}$ increases with increasing energy.

\end{enumerate}

\subsection{Electron spectra}
\label{sec:electron_spectra}

Once we have the density of electrons in all points of the space, we compute the quantity of electrons that are inside the volume given by the line of sight from the earth to the source. 





\begin{figure}
\begin{center}
\includegraphics[width=0.48\textwidth]{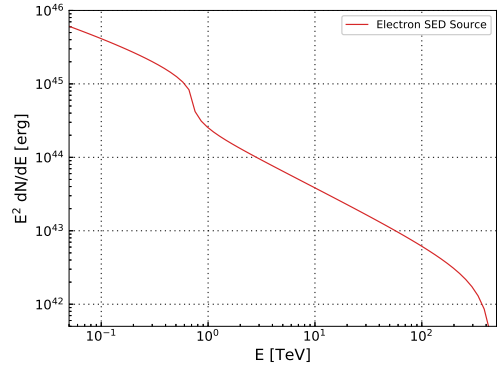}
\caption{Spectral energy distribution of electrons from the source. }
\label{fig:electron_spectra}
\end{center}
\end{figure}

The selected size for the source is $\theta=5^\circ$. We compute the differential energy of the electrons contained inside the source's volume. To do it, we integrate over a sphere centered at the pulsar position with its boundaries limited by the size of the cone. The result of this integration is the differential energy spectrum of all the electrons that are producing the $\gamma$-ray emission. The spectral energy distribution of these electrons is shown on the right panel of Figure \ref{fig:electron_spectra}. The kink present at $E\sim1$ TeV represents the transition from the energy where electrons are not cooled to that where they are cooled. The difference in spectral indices between the these two regions has the same origin. If we computes the total differential output of electrons in the case of no cooling and slow diffusion, situation where all the injected electrons should be contained inside the integrated volume, we recovered the total energy of the injected electrons.


\subsection{Gamma-ray spectra}
\begin{figure}
\begin{center}
\includegraphics[width=0.48\textwidth]{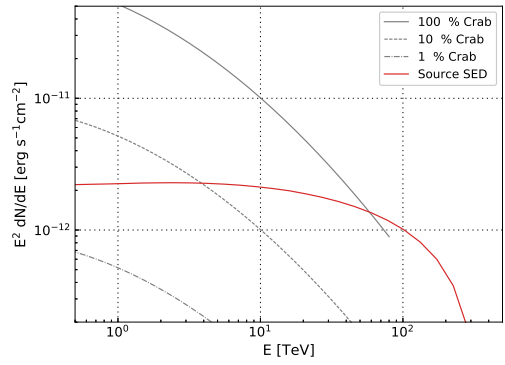}
\caption{Gamma-ray spectra for a source of $\theta=5^\circ$. Gray lines showing the Crab flux are also included.}
\label{fig:gamma_spectra}
\end{center}
\end{figure}

The gamma-ray spectrum at TeV energies is produced by IC up-scattering of ambient photons, mainly CMB at multi-TeV energies. As mentioned in Section \ref{sec:inverse_compton}, we use Eq. 2.48 from \cite{Blumenthal70} to calculate the IC emissivity of each of the samples shown in Figure \ref{fig:electron_spectra}. The results are shown on Figure \ref{fig:gamma_spectra}.

\subsection{Modeling $e^\pm$ Propagation to Earth}

\begin{figure}
\begin{center}
\includegraphics[width=0.48\textwidth]{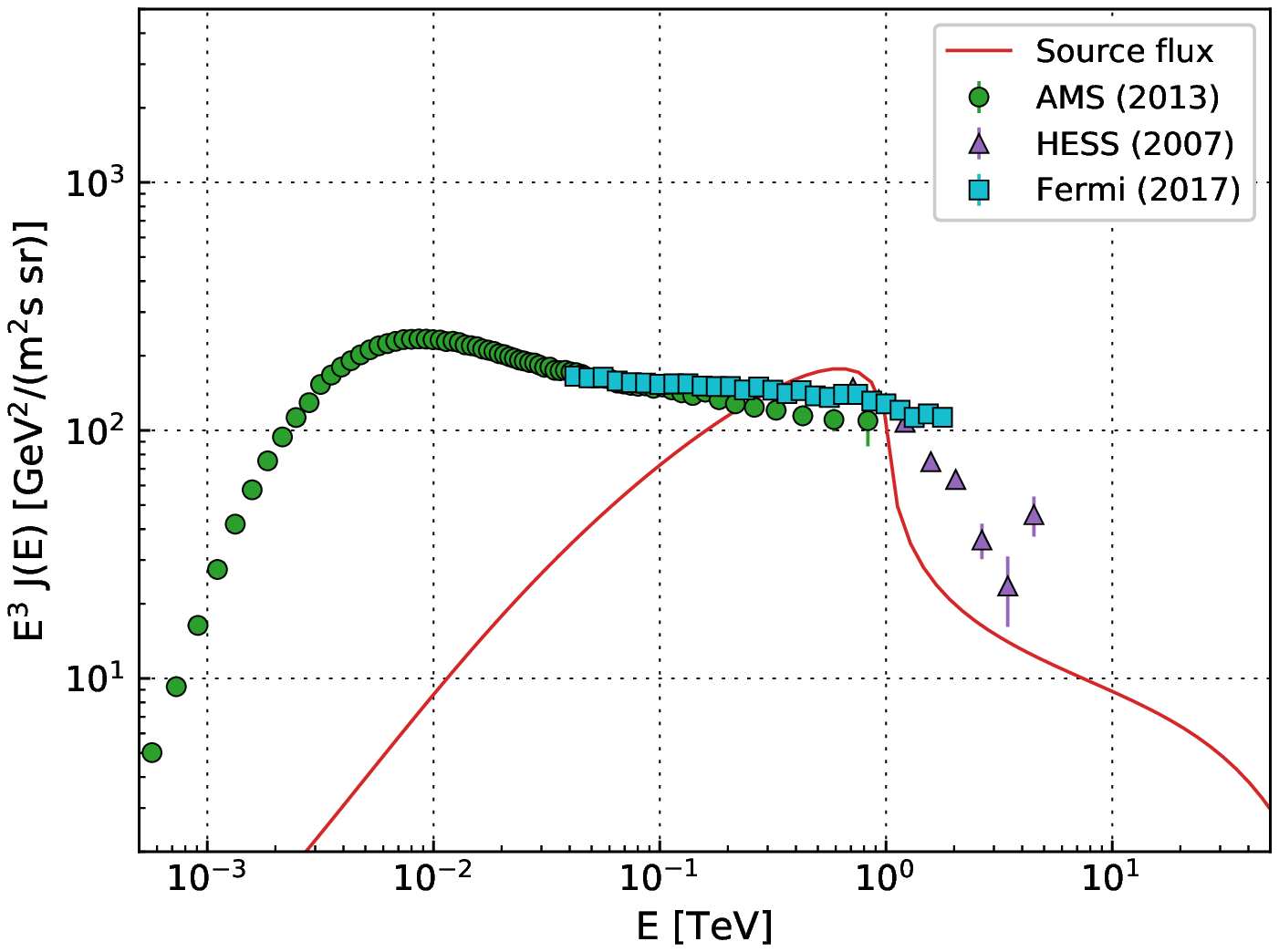}
\includegraphics[width=0.48\textwidth]{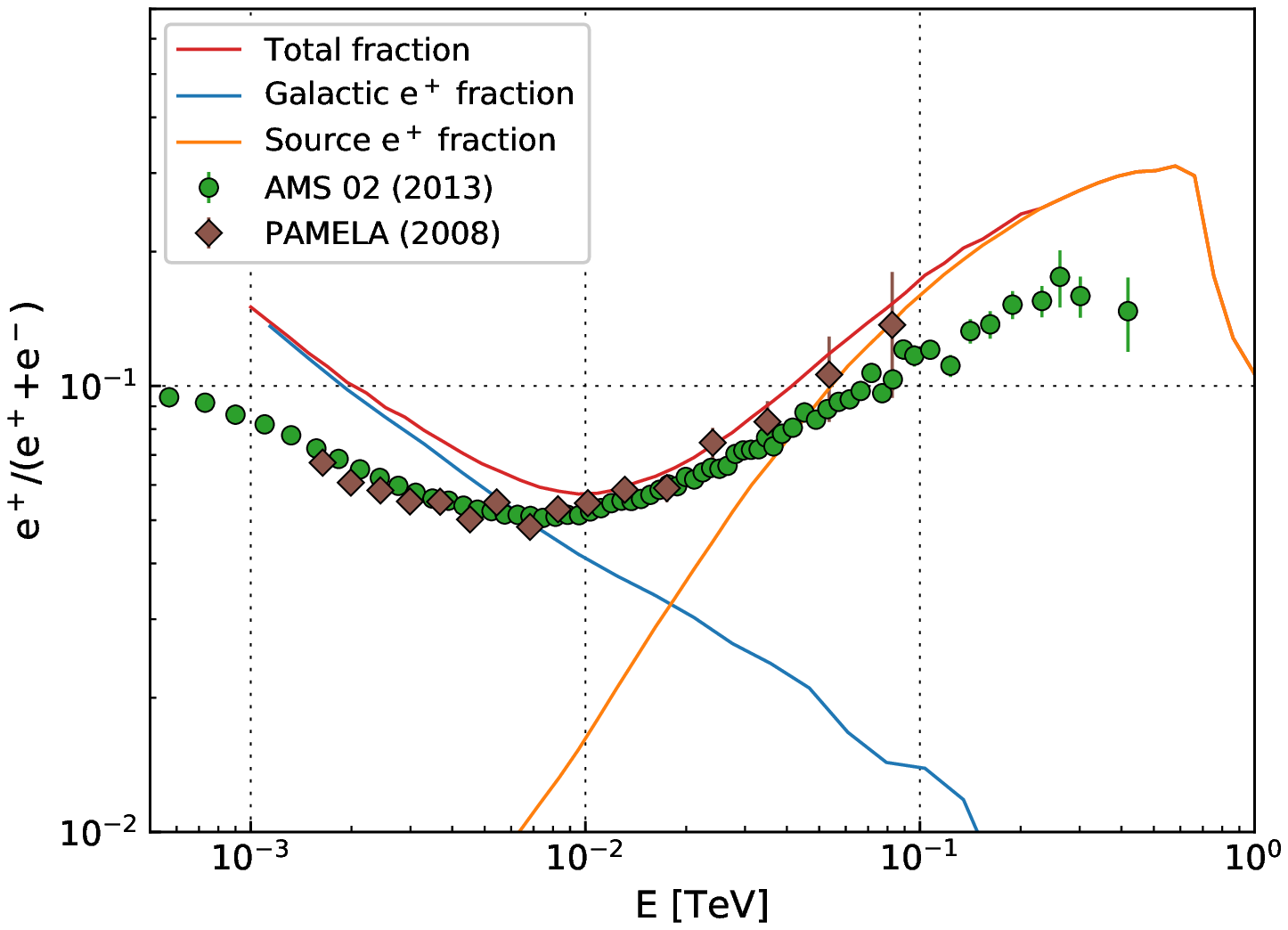}
\caption{All-electron flux at the Earth (left panel) and positron fraction (right panel).}
\label{fig:flux_fraction_Earth}
\end{center}
\end{figure}

To compute the electron and positron flux provided by a given source at the Earth, we use Eq. \ref{eq:flux_earth}. For the fraction of positrons that this source contributes to, we use Eq. \ref{eq:fraction_earth}. The results are shown on Figure \ref{fig:flux_fraction_Earth}.

\section{Conclusion}
We developed a code to calculate the diffusion of electrons and positrons from point-like sources to the Earth. With the code we can calculate the distribution of electrons and positrons produced by a central source, the $\gamma$-ray spectrum produced by this source and the electron and positron flux produced by these source at the Earth, as well as the positron fraction.

The code can be found in the github repository https://github.com/rlopezcoto/EDGE

\section*{Acknowledgements}
The authors would like to thank the HAWC collaboration for useful discussions during the development of the electron diffusion code.

\bibliographystyle{JHEP}
\bibliography{references} 

\end{document}